\documentclass[10pt]{article}

\usepackage[utf8]{inputenc}
\usepackage[english]{babel}
\usepackage[left=2cm,right=2cm,top=2cm,bottom=2cm]{geometry}
\usepackage{graphicx}

\usepackage{amsmath}
\usepackage{amsfonts}
\usepackage{amssymb}

\usepackage{cite}

\usepackage{amsthm}

\newtheorem{definition}{Definition}[section]

\usepackage[acronym]{glossaries}
\newacronym{mcs}{MCS}{Maximum Common Subgraph}
\newacronym{mwcs}{MWCS}{Maximum Weighted Common Subgraph}
\newacronym{lmwcs}{LMWCS}{Labeled Maximum Weighted Common Subgraph}
\newacronym{lmcs}{LMCS}{Labeled Maximum Common Subgraph}
\newacronym{qubo}{QUBO}{Quadratic Unconstrained Binary Optimization}
\newacronym{lmwis}{LMWIS}{labeled maximum weighted independent set}
\newacronym{mis}{MIS}{Maximum Independent Set}

\usepackage{geometry}                
\usepackage[parfill]{parskip}
\usepackage{graphicx}
\usepackage{epstopdf}
\usepackage{xargs}
\usepackage{ifthen}
\usepackage{extpfeil}
\usepackage{tikz}
\usepackage{dsfont}
\usepackage{pgf}
\usepackage{stmaryrd}
\usepackage[verbose]{wrapfig}
\usepackage{xcolor}
\usepackage{url}
\usetikzlibrary{arrows}
\usepackage{amssymb}
\usepackage{amsmath,multirow}
\usepackage{caption}
\usetikzlibrary{positioning,shapes.geometric, arrows}

\usepackage{caption}
\usepackage{pdflscape}
\usepackage{adjustbox}

\usepackage[T1]{fontenc}
\usepackage[utf8]{inputenc}
\usepackage{authblk}
\title{A Novel Graph-based Approach for Determining Molecular Similarity}
\author[1]{Maritza Hernandez\thanks{maritza.hernandez@1qbit.com}}
\author[1,2]{Arman Zaribafiyan}
\author[1]{Maliheh Aramon}
\author[3]{Mohammad Naghibi}
\affil[1]{{\textit{1QB Information Technologies (1QBit), Vancouver, BC V6B 4W4, Canada}}}
\affil[2]{{\textit{Department of Electrical and Computer Engineering, University of British Columbia, Vancouver, BC, V6T 1Z4, Canada}}}
\affil[3]{{\textit{Department of Computer Science, University of Calgary, Calgary, AB T2N 1N4, Canada}}}
\date{}
\begin{document}
\maketitle

\begin{abstract}
In this paper, we tackle the problem of measuring similarity among graphs that represent real objects 
with noisy data. To account for noise, we relax the definition of similarity using the maximum weighted co-$k$-plex 
relaxation method, which allows dissimilarities among graphs up to a predetermined level. 
We then formulate the problem as a novel quadratic unconstrained binary optimization problem that can be solved 
by a quantum annealer. The context of our study is molecular similarity where the presence of noise might 
be due to regular errors in measuring molecular features. We develop a similarity 
measure and use it to predict the mutagenicity of a molecule. Our results indicate that 
the relaxed similarity measure, designed to accommodate the regular errors, yields a higher 
prediction accuracy than the measure that ignores the noise.
\end{abstract}


\section{Introduction}
\label{sec:introduction}

The accumulation of data in today's digital world is growing exponentially, affecting 
various fields, such as physics, decision sciences, astronomy, and biology. \emph{The Economist} estimates 
that by the year 2020, the amount of data in the world will be seven times greater than the data that 
existed in 2014 \cite{BBCmagazine}. The revolutionary impact of ``big data'' in biology has been tremendous, 
transforming it into data-rich field. The majority of the new data, including data sets of protein structures, DNA sequences, 
and gene expressions are complex, calling for the development of new analytical tools and algorithms to translate the data into valuable insights. 
The first challenge then lies in the efficient representation of biological data objects. Graphs have shown promising potential in representing data objects 
including DNA \cite{Kehr14} and small molecules \cite{Raymond02} due to their high abstractional capacity \cite{Eshera86}. 
The problem of biological data analysis is then reduced to graph analysis. The goal is to search, visualize, and infer 
new correlations among objects to produce healthier foods, prevent diseases, and discover individually customized drugs. 
Therefore, meaningful metrics are required to determine similarity and dissimilarity among graphs.

Graph similarity methods are developed primarily based on the property 
 of isomorphism in the graph theory literature \cite{Zager}. A graph $G_1 = (V_1,E_1)$ is called isomorphic 
 to a graph $G_2 = (V_2,E_2)$ if there is a bijection between the vertex sets $V_1$ and  $V_2$ such that 
 there exists a mapping between the adjacent pairs of vertices of $G_1$ and $G_2$. This is an ``edge-preserving" 
 bijection, consistent with the general notion of ``structure-preserving". The main drawback of isomorphism in 
 comparing real objects is its restrictive binary nature---it does not consider partial similarities among graphs. Two graphs are 
 evaluated as being either exactly similar up to label permutation or not similar at all. Therefore, practical variations 
 of isomorphism have been developed, including edit distance and maximum common subgraph (MCS) \cite{Bunke00}. 
 The latter has a structure-preserving feature similar to isomorphism. The MCS of two graphs $G_1$ and $G_2$ is 
 the largest subgraph of $G_1$ that is isomorphic to a subgraph of $G_2$. 
 
In real-world applications, particularly in biology, graphs of objects carry different types of information whose level of importance may 
also differ. To account for real-world properties, the MCS problem has been generalized to the labelled 
maximum weighted common subgraph (LMWCS) problem, where vertices and/or edges 
are associated with multiple labels, storing various information. For example, in a molecular similarity 
application, a molecular graph is built where vertices and edges usually represent atoms and atomic bonds, respectively. 
The vertices may have labels for atomic numbers and formal charges, and the labels of the edges may indicate 
the bond types. It has been shown that the MCS and LMWCS problems are equivalent to another well-known 
problem---the maximum independent set (MIS) of a third graph, which can be induced from the graphs 
being compared \cite{Khazm07,Balasundaram13}. The vertices and edges of the third graph, respectively, 
represent possible mappings and the conflicts between them. The goal of the MIS problem is to find the largest set of 
 vertices such that there is no edge between all selected pairs, forming the largest conflict-free mapping.

Representing the MCS problem as the MIS problem makes it easier to further relax the definition of similarity 
where the data is noisy or distorted. In such contexts, the definition of similarity based on 
exact subgraph matching might be prohibitive, excluding useful information. For example, in three-dimensional 
graph representation of molecules, the position of each atom is obtained relative to the other atoms by solving 
an optimization problem heuristically. Therefore, the graph representation is partially accurate. A more flexible similarity 
measure can compensate for the effect of noise in the data, considering as similar substructures with 
conflicts up to a certain tolerable threshold. Even if there is no perturbation in the data, 
relaxing the definition of similarity can be beneficial. For example, to compare a newly sequenced 
genome with a reference genome, relaxing the definition of similarity allows the regular variations that occur within 
populations to be accommodated. There are different relaxations of the exact 
subgraph matching requirement in the literature \cite{k_plex_social}. One of the relaxations is known as the 
maximum co-$k$-plex problem, where the goal is to find the largest set of vertices in the graph such that 
each vertex has at most $k-1$ edges connecting it to the other vertices. It is clear that the maximum co-$1$-plex problem 
is the MIS problem \cite{Balasundaram13}. 

The majority of the similarity methods discussed above, including the MIS and maximum co-$k$-plex problems, 
are in general NP-hard, having exponentially increasing computational complexity due to the combinatorial nature 
of graphs \cite{Garey79,Downey95}. However, since a given graph representation can be used to model 
a wide range of problems, including image processing and DNA sequencing, graph-based algorithms are practical. As 
a result, there have been efforts to find heuristics with lower complexity, develop approximation methods, restrict graph 
structures, or use parallel implementation strategies to solve graph similarity problems \cite{Wang97,Gupta96,Kollias14}. These methods, 
though designed to have polynomially bounded complexity, fail to find the optimal solutions. It is, furthermore, 
challenging to evaluate their performance, gaining insights on how much we lose by not using the exact algorithms in the
worst case. 

Recent advances in the ability to utilize quantum mechanical effects in computation have sparked interest in the 
research community that quantum computing could provide a speed-up in solving classical 
NP-hard optimization problems \cite{Santoro06,Farhi01}. Quantum annealing is one approach to harnessing 
quantum effects in searching the energy landscapes of optimization problems. \emph{D-Wave Systems} has developed 
a physical realization of a quantum annealer \cite{Johnson11}, which has been shown to exhibit quantum effects 
such as tunnelling  \cite{Boixo15} and entanglement  \cite{Lanting14}. This device finds 
the minimum value (i.e., ground state) for a special class of objective functions (i.e., Ising objective functions). 
The class of objective functions native to the hardware can be translated to quadratic functions of binary 
variables, and since the hardware cannot encode constraints, the device is a quadratic unconstrained binary 
optimization (QUBO) problem solver. As a result, there has been work in recent years to formulate various 
constrained optimization problems as QUBO problems \cite{Rieffel15,Venturelli15,Rosenberg15}. 

One of the main contributions of this paper is to propose a novel QUBO problem formulation for the 
maximum weighted co-$k$-plex problem, measuring similarity among graphs. It is worth mentioning that our 
formulation can be solved by a quantum annealer; however, it is not our aim in this paper to study 
the annealer's performance. Our focus is to investigate the performance of the QUBO-based similarity 
measure in the context of a real molecular similarity problem. To accurately evaluate the performance of the QUBO-based measure over 
 existing molecular similarity measures, we use an exhaustive solver to find the optimal similarity values among 
graphs.  

The rest of the paper is organized as follows. In Section \ref{graph_similarity_methodology}, we review the 
existing QUBO problem formulation for the MIS problem and present a novel formulation for the maximum 
weighted co-$k$-plex problem, measuring the similarity among graphs. In Section \ref{molecularsimilarity}, 
we use the new model in the context of molecular similarity where the reduction of molecules to graphs and 
the adaptation of the novel QUBO-based formulation to molecular similarity are explained. Finally, in 
Section \ref{sec:Exp_result}, we use a machine learning approach on real molecular data sets to evaluate 
the performance of our similarity measure in predicting mutagenicity. 

\section{A QUBO-based Graph Similarity Measure} 
\label{graph_similarity_methodology}

The MIS problem has been formulated as a QUBO problem \cite{Boros02,DW_Embedding_1}; however, to our knowledge 
there is no QUBO problem formulation that considers the relaxation of the exact subgraph matching. 
In this section, we first review the existing formulation of the MIS problem. We then introduce 
our novel formulation for the maximum weighted co-$k$-plex problem and discuss its generalization to similarity problems among multiple graphs.  
Finally, we explain how to quantify the similarity between the labelled graphs given the solution to the QUBO problem formulation. 

\subsection{QUBO Problem Formulation of MIS} 
\label{subsec:LWMCS_formulation}

As discussed, there is a mapping between the MIS and LMWCS problems. Here, we review the 
QUBO problem formulation for the latter problem and then show how it can be converted into a 
QUBO problem for the MIS problem.

Let us consider two arbitrary graphs $G_1=(V_1,E_1)$ and $G_2=(V_2,E_2)$. To formulate the LMWCS problem 
as a QUBO problem, one binary variable $b_{ij}$, defined below, is assigned to each possible pair $(i, j) \in V_1 \times V_2$. The 
weight of each pair is denoted by $w_{ij}$. A $0,1$ configuration of binary variables $b_{ij}$ depicts a common 
subgraph between $G_1$ and $G_2$. The vertex pairs that are assigned to 1 form the common subgraph. 
In other words, they are in the mapping. The goal is to find a configuration that maximizes the weight of 
the common subgraph, satisfying both the bijection and the user-defined requirements. 

Formally, let
  \begin{equation}
    b_{ij} =
    \begin{cases}
      1  & \text{if}\  (i,j) \in \text{mapping}, \\ \nonumber
      0  & \text{otherwise}.
    \end{cases}
  \end{equation}

The product set $V_1 \times V_2$ does not necessarily include all possible pairings between two vertex sets of the graphs---
the vertex labels can exclude a specific pair from the mapping. For example, in a biology application, 
only the vertices with the same atomic number might be paired together.

Let us further define two sets to impose, respectively, the bijection and user-defined constraints on the mapping: 
\begin{align}
C&= \big\{ \big((i,j),(m,n)\big) \big| ~i=m \vee j=n \big \},  \notag \\
C^*&= \big \{ \big( (i,j),(m,n) \big) \big |~ b_{ij}b_{mn}=0 \:\: \text{is a user-defined constraint} \big \}. \nonumber 
\end{align}

The QUBO problem formulation of the LMWCS problem is equivalent to the following formulation for all 
$a_{(i,j),(m,n)} > \mbox{min}\{w_{ij},w_{mn} \}$ \cite{DW_Embedding_1}:
\begin{equation}
\label{eq:MCS_QUBO}
\max  \Bigg( \sum_{(i,j) \in V_1 \times V_2} w_{ij}b_{ij} - \sum_{\big((i,j),(m,n)\big) \in C \cup C^*} a_{(i,j),(m,n)}b_{ij}b_{mn} \Bigg).
\end{equation}

The first expression in the above formulation maximizes the weights of the selected 
pairs of vertices and the second expression penalizes the infeasible assignments due either to the 
bijection requirement $C$ or the user-defined requirements $C^*$. 
An example of the latter requirements is when the user is interested in finding the maximum 
common clique between two graphs. The set $C^*$ then includes 
$\big( (i,j),(m,n) \big)$ if $(i,m) \notin E_1$ or $(j,n) \notin E_2$. It is worth mentioning that 
the degree of freedom represented as set $C^*$ allows the enforcement of customized 
structure restrictions on the resulting MCS.

To convert Formulation \eqref{eq:MCS_QUBO} to an MIS formulation, we introduce a third graph called a conflict graph. 
Formally, let $G_{c}=(V_c, E_c)$ be the conflict graph of graphs $G_1$ and 
$G_2$, where $V_c \subset V_1 \times V_2$ and $E_c = C \cup 
C^*$. The vertices of the conflict graph are 2-tuples of the labelled graph 
vertices. All possible 2-tuples are not necessarily the vertices of the conflict graph---some 
pairs might be excluded by the user based on predetermined criteria. The edges of the conflict graph 
represent both the bijection and the user-defined requirements, defined by $C$ and $C^*$, respectively. 
That is, there is an edge between two vertices $s=(i,j)$ and $l=(m,n)$ in the conflict graph 
if and only if the tuple of the corresponding pairings $\big((i,j),(m,n)\big)$ exists 
in $C \cup C^*$. The edges are called \emph{conflicts} 
because they identify the pairs that cannot be present in the common subgraph simultaneously. Therefore, 
it is straightforward to see that the MIS of the conflict graph $G_c$ corresponds to the MCS of 
graphs $G_1$ and $G_2$.

The QUBO problem formulation for the MIS problem of the conflict graph is \begin{align}
\label{eq:MIS_QUBO}
\mbox{max~~} \Bigg( & \sum_{\substack{s \in V_c}} w_sx_s -  \sum_{\substack{(s,l) \in E_c}} a_{sl} x_s x_l \Bigg) ,
\end{align}
\noindent where $x_s$ is a binary variable that is equal to 1 if the vertex $s$ is included in the independent set and 0 otherwise, $w_s=w_{ij}$, and $a_{sl} > \mbox{min} \{w_s,w_l\}$.

It can be demonstrated that both QUBO problem Formulations \eqref{eq:MCS_QUBO} and \eqref{eq:MIS_QUBO} 
are equivalent by a simple notation transformation.

\subsection{QUBO Problem Formulation of Maximum Weighted Co-$\textbf{\emph{k}}$-plex}
\label{subsec:cokplex_QUBO}

A co-$k$-plex of a graph $G$ is a subgraph of $G$ in which each 
vertex has a degree of at most $k-1$. Therefore, the maximum weighted co-$k$-plex of 
the conflict graph $G_c$ corresponds to the maximum weighted common subgraph of $G_1$ and $G_2$, 
where each possible pairing can violate at most $k-1$ constraints, either bijection or user-defined. 
To generalize QUBO problem Formulation \eqref{eq:MIS_QUBO}, let us first define a star graph.
\\
\begin{definition}
\label{stardef}
A graph $S^k=(V, E)$ is a star graph of size $k$ if it is a tree with $k+1$ vertices and one vertex of degree $k$. 
\end{definition}

Based on the co-$k$-plex definition, we do not penalize the conflict edges. Each vertex in the conflict 
graph can have up to $k-1$ edges. Therefore, we only penalize in situations where a subset of the vertices 
induces a subgraph in which there is one vertex with degree greater than $k-1$. In other words, 
we penalize all subsets of vertices whose induced subgraph forms a star graph of size $k$.

Further, let us define the binary parameter $\mathcal{A}_{v_1, \ldots, v_{k+1}}$ as
\begin{align}
\mathcal{A}_{v_1, \ldots, v_{k+1}} = \begin{cases}
1   & \text{if~} \{v_1, \ldots, v_{k+1} \} \text{~induces~} \,\, S^{k}, \\
0   & \text{otherwise}, \notag
\end{cases}
\end{align}
where $v_1, \ldots, v_{k+1}$ are the vertices of the conflict graph $G_c$.

The QUBO problem formulation of the maximum co-$k$-plex is
\begin{align}
\label{eq:co-k-plex_QUBO}
\mbox{max~~} \Bigg(  & \sum_{\substack{v_i \in V_c}} w_{v_i} x_{v_i} - \big(\sum_{\substack{(v_1,\ldots,v_{k+1})}} a_{v_1,\ldots,v_{k+1}} 
                                           \mathcal{A}_{v_1,\ldots,v_{k+1}} \prod\limits_{i=1}^{k+1}x_{v_i} \big) \Bigg) ,
\end{align}
where 
$x_{v_i}$ is a binary variable equal to 1 if the vertex $v_i$ is included in the MIS or 0 otherwise, and $a_{v_1, \ldots, v_{k+1}} > \mbox{min} \{w_{v_1},\ldots,w_{v_{k+1}}\}$. 
The $k$ parameter is a tunable parameter that should be determined by the user. 

The above formulation allows both bijection and user-defined conflicts to be present in the MIS. However, the formulation can be modified as below to exclude any bijection conflicts 
in the independent set of the conflict graph for $k \geq 2$:
\begin{align}
\label{eq:co-k-plex_QUBO_nobijection}
\mbox{max~~} \Bigg(  & \sum_{\substack{v_i \in V_c}} w_{v_i} x_{v_i} - \sum_{\substack{(v_1,v_2) \in \mathcal{C}}} a_{v_1,v_2} x_{v_1} x_{v_2}  
                                                                                                  - \big (\sum_{\substack{(v_1,\ldots,v_{k+1})}} a_{v_1,\ldots,v_{k+1}} 
                                                                                                     \mathcal{A}_{v_1,\ldots,v_{k+1}} \prod\limits_{i=1}^{k+1}x_{v_i} \big)\Bigg).
\end{align}
 
As mentioned in Section \ref{subsec:LWMCS_formulation}, the set $C$ includes 
pairs of vertex sets $V_1$ and $V_2$ (alternatively, the vertices of the conflict graph) that cannot 
be simultaneously present in the common subgraph (alternatively, the independent set of the conflict graph). 
Therefore, the second expression in Formulation \eqref{eq:co-k-plex_QUBO_nobijection} penalizes 
the conflict edges due to the bijection requirement. Note that the definition of $\mathcal{A}_{v_1,\ldots,v_{k+1}}$ also 
slightly differs in this case: it equals 1 if vertices $v_1, \ldots, v_{k+1}$ induce a star graph of $S^k$ 
whose edges are only the elements of set $C^*$, that is, the user-defined constraints.

The objective functions of both Formulations \eqref{eq:co-k-plex_QUBO} and 
\eqref{eq:co-k-plex_QUBO_nobijection} are higher-order polynomials. There are several algorithms in the 
literature that map higher-order polynomials to quadratic polynomials \cite[and references therein]{hobo2qubo}.

\paragraph{Multiple Graph Similarity} One of the main advantages of our conflict graph-based QUBO problem formulation 
is that it can easily be extended to measure similarity among multiple 
graphs. An example of a multiple graph similarity problem in molecular biology is the 
comparison of an unknown molecule with several sets of molecules to find the set that is the most similar. 
Other examples are human face recognition, where the goal is to find an image in a database that best matches several 
images of the same person taken from different angles \cite{Bengoetxea02}, and graph prototype 
construction in clustering and classification contexts, in order to best represent all elements of a set of labelled 
graphs \cite{Ribalta12}.

Let us formally assume that there are $n$ labelled graphs $G_i=(V_i,E_i)$, where $i=1,2,\ldots, n$, to compare. 
An obvious strategy is to find the maximum weighted common subgraph of graph pair $G_i$ and $G_{i+1}$ 
for $i=1, ..., n-1$ and then infer the global maximum weighted common subgraph. Following this 
strategy, we need to build $n-1$ conflict graphs and solve their corresponding NP-hard QUBO problem 
formulations. However, we can generalize our graph similarity approach to find the 
maximum common subgraph among a set of graphs by solving only \emph{one} QUBO problem. 

The core idea of comparing the $n$ labelled graphs is the same as before: building the conflict graph and 
translating it into a QUBO problem formulation. The vertex set of the conflict graph of $n$ graphs, $V_c$, 
is a subset of the product of $V_1 \times V_2 \times \cdots \times V_n$. Therefore, each 
vertex of the conflict graph is an $n$-tuple of individual graph vertices. The tuples that 
do not satisfy predefined compatibility criteria are excluded. There is an edge between two 
vertices in the conflict graph if there is any reason imposing that the two vertices cannot 
be present in the similarity set at the same time. The maximum weighted 
independent set problem of the conflict graph can then be formulated as a QUBO problem following 
the same procedure as already explained.


\subsection{Similarity Measure}
\label{Similarity_Measure}

The solution to the QUBO problem formulation presented in the previous section is a binary vector with 
size $|V_c|$. The non-zeros in the solution vector identify the pair of vertices present in the maximum 
weighted independent set of the conflict graph. To measure the similarity between graphs, 
there exists a large number of similarity metrics in the literature, represented mainly in terms 
of subset relations  \cite{Baum07,Raymond02}. In this paper, we 
use the metric 
\begin{equation}
\mathcal{S}(G_1,G_2) = \delta \max \Bigg\{ \frac{|V^1_c|}{|V_1|}, \frac{|V^2_c|}{|V_2|} \Bigg\} + (1-\delta) \min \Bigg\{ \frac{|V^1_c|}{|V_1|}, \frac{|V^2_c|}{|V_2|}  \Bigg \}, \quad \delta \in [0,1],
\label{simEq}
\end{equation}
where $|V^1_c|$ and $|V^2_c|$, respectively, denote the number of distinct vertices of 
$G_1$ and $G_2$ in the maximum weighted independent set of the conflict graph. This metric 
quantifies the contribution of each graph to the MIS. Our metric is 
the convex combination of two existing similarity measures---Bunk and Shearer, and Asymmetric \cite{Baum07}--- 
providing the user with more flexibility (see Section \ref{sec:Exp_result} for more details). 

In the next section, we utilize our QUBO-based graph similarity measure in the context of molecular similarity.

\section{Molecular Similarity}
\label{molecularsimilarity}

Molecular similarity methods play an important role in many aspects of chemical and 
pharmaceutical research, including biological activity prediction \cite{Waterbeemd03}, 
protein-ligand docking \cite{Kuhl84}, and database searching \cite{Willett99}. Determining similarity between compounds is practical due to the 
\emph{similar property principle} \cite{principle}, which states that structurally similar molecules are expected 
to display similar properties. Therefore, various aims in biological research can be accomplished by 
determining structural similarity between molecules. Similarity, however, is a complicated concept to evaluate 
since it is not a measurable property of the molecules themselves. Indeed, it may have 
different interpretations depending on external criteria. This ambiguity increases the difficulty of 
developing an accurate computational method for determining molecular similarity. 

To build a measure of molecular similarity, three basic components are required: a molecular 
representation that not only encodes relevant molecular features but also contains their 
associated weights, a method that compares molecular representations, and a function that evaluates 
their similarity. Many similarity measures have been proposed in the literature which can be 
categorized into two classes. The first class of measures uses the intuitive representation of the 
molecular atom-bond structure as a graph. Comparison of molecules based on graph 
representations is often accomplished using graph similarity techniques. As mentioned in 
Section \ref{sec:introduction}, the computational cost of graph similarity 
techniques has resulted in the development of algorithms for specific graph structures. 
These approaches present limitations on the types of objects that can be used for comparison. 
However, the QUBO-based measure presented in this paper is applicable to any arbitrary graph. 

The second class of measures uses a vector-based representation 
called fingerprint, a conventional concept in chemical informatics and related fields. 
Fingerprints are binary vectors that indicate the presence or absence of certain 
features of a molecule. Each bit can be either 1 or 0, representing whether or not the 
molecule contains an associated feature. Molecules modelled as fingerprints are compared 
using vector-based comparison measures such as Tanimoto, Cosine,  Dice, and Euclidean 
distance \cite{Baum07}. Although fingerprints are easy to use and their pairwise comparisons are 
computationally efficient, they have certain drawbacks. For instance, fingerprints cannot be used to  
assess for certain whether a particular pattern is present or not in a molecular graph  \cite{Baum07}. When a bit in the 
fingerprint is set to $1$ for some pattern, it means that the pattern is present in the molecule 
only with some probability. Furthermore, fingerprint-based approaches do not usually consider 
underlying information about the molecular topology.

In the next section, we discuss how our graph similarity method, described in 
Section \ref{graph_similarity_methodology}, can be applied to the molecular similarity problem.

\subsection{Graph-based Molecular Similarity}
\label{graphbasedmolecularsimilarity}

To measure similarity between molecules, we discuss how to represent molecules as graphs 
and then build a conflict graph. Given the conflict graph, it is then straightforward to formulate the 
QUBO problem and quantify the similarity as described in Sections \ref{subsec:cokplex_QUBO} and \ref{Similarity_Measure}.

\paragraph {Modelling Molecules as Graphs} A direct graph representation of the structural formula 
of a chemical compound is often referred to as a molecular or chemical graph. In general, 
a molecular graph-based representation can be categorized into two classes based on 
the structural dimensionality of the molecule, that is, two-dimensional (2D) and three-dimensional (3D). 
In the former representation, each vertex is associated with one atom and there 
is an edge between two vertices indicating their chemical bond. The latter representation
includes 3D structural information. For example, in the 3D molecular graph introduced by 
Raymond and Willett \cite{Raymond02}, there is an edge between each pair of vertices indicating the geometrical distance 
between the two corresponding atoms. 

\begin{figure}[h]
\centering
\includegraphics[scale=0.6]{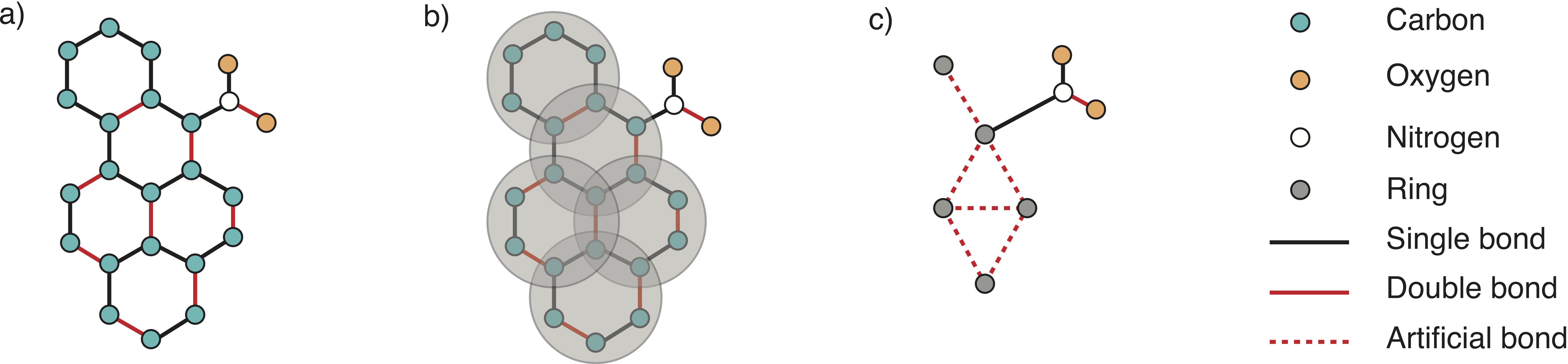}
\caption{Steps for modelling molecules as graphs: the first step is to a) build a basic structural representation of a
molecule, then b) identify ring structures, which subsequently are added as a vertex to the reduced graph as shown
in c).}
\label{fig:MR}
\end{figure}

Since a graph is an abstract mathematical object, vertices and edges do not need to correlate 
with atoms and bonds directly. Hence, vertices could represent a group of atoms, given 
some meaningful criteria. Accordingly, a reduced representation of the chemical 
structures reduces the size of the graphs being compared and thus increases the 
effectiveness of the algorithms to solve the MIS (or its relaxation) problem. In this paper, we build a labelled 
reduced graph representation of a molecule. The steps of this reduction process are 
illustrated in Figure \ref{fig:MR}. The first step is to construct the atom-bond graph. 
We then identify ring structures of the molecules and reduce each 
corresponding ring's set of vertices to a single vertex. If two vertices representing rings 
share one or more atoms, an edge between these vertices is then added and 
labelled as \textit{artificial}, emphasizing  that it does not represent a natural chemical bond. 
Furthermore, vertices and edges of our graph representation are labelled with 
various information pertaining to atoms and bonds, respectively. 

Formally, the labelled reduced chemical graph can be written as 
$G=(V,E,\mathcal{L}_{V},\mathcal{L}_{E})$, where $V$ is the set of vertices 
and $E=V \times V$ is the set of edges. A set of labels,  $\mathcal{L}_{V} : V \rightarrow \{ l_{1}, \ldots , l_{8} \} $, 
is assigned to each vertex. Each label represents a specific property of the atom or ring. 
The label set contains the atomic number, the number of implicit hydrogen 
bonds\footnote{Explicit hydrogen atoms are added as vertices to the graph.} associated with the 
atom (or ring), the formal charge, the degree of the node distinguishing  
covalent bond order (single, double, or triple), the set of covalent bond order between 
atoms in a ring, a weight denoting the number of atoms reduced to one 
vertex, and a 3D vector indicating the position of the atom or the 
geometrical centre of the ring. Likewise, a label $ l_{e}$ is assigned  to each 
edge ($\mathcal{L}_{E} : E \rightarrow l_{e}$), representing whether the edge is 
an artificial link, or a single, double, or triple covalent bond.

\paragraph{Building a Conflict Graph}
The conflict graph $G_{c}=(V_{c},E_{c})$ of two chemical graphs $G=(V,E)$ and $G'=(V',E')$ is 
defined, according to Section \ref{subsec:LWMCS_formulation}, by its vertex 
set $V_{c} \subseteq V \times V' $ and edge set $E_{c} \subseteq V_{c} \times V_{c}$ as:
\begin{eqnarray}
V_{c} &=& \{ (v_{i},v'_{j}) \, |  \, \mathcal{L}_{V}(v_{i}) = \mathcal{L}_{V}(v'_{j}) \}, \\ \label{Vc}
E_{c} &=& \{ ((v_{i},v'_{j}),(v_{k},v'_{l}) ) \, | \, i = k  \vee \,  j = l \vee \, \mathcal{L}_{E}(v_{i},v_{k}) \neq \mathcal{L}_{E}(v'_{j},v'_{l})  \} \label{Ec},
\end{eqnarray}
where $\mathcal{L}_{V}$ is the set of labels assigned to the vertices as specified 
above. The vertex set $V_{c}$ includes a pair of vertices from the chemical graphs under 
comparison if they have matching labels. It should be noted that each of the labels can be 
set as a default or optional parameter according to some similarity criteria for the specific application. 
Here, we assume that the atomic number, the weight, and the number of hydrogen bonds associated 
to atoms are matching properties set by default, while the number of hydrogen bonds 
associated to rings (RH), the bond order within a ring (RB), the formal charge (FC), and the degree of 
the nodes (DN) are optional parameters. The first two conditions associated with the creation of edges 
in \eqref{Ec} imply that a conflict edge between two vertices $(v_{i},v'_{j})$ and $(v_{k},v'_{l})$ is added 
if one of the vertices from the chemical graphs has been matched twice, that is, if the bijection 
requirement is violated. The third condition in \eqref{Ec} indicates that a conflict edge is added if 
there is a mismatch between the edge labels of the two chemical graphs. The set of conflict 
edges can be extended to include 3D molecular information. Since molecular compounds 
are inherently 3D, it is expected that the incorporation of 3D information would be preferable 
over 2D. To include 3D information, we consider the geometrical distance between a pair of atoms. 
That is, an edge between two vertices in the conflict graph (e.g., between $(v_{i},v'_{j})$ and $(v_{k},v'_{l})$) 
is added if the distance between the vertices $v_{i}$ and $v_{k}$ of the graph $G$ is not 
comparable to the distance between vertices $v'_{j}$ and $v'_{l}$ of the graph $G'$. 
To determine if two distances are comparable, we introduce a user-defined threshold 
distance $d_{t}$. Thus, a new condition can be added to \eqref{Ec},
\begin{equation}
E_{c} = \Big\{ ((v_{i},v'_{j}),(v_{k},v'_{l}) ) \, \Big | \,  | d(v_{i},v_{j}) {-} d(v'_{j},v'_{l}) | \, >  \, d_{t}  \Big \} ,
\end{equation}
where $d(v_{i},v_{j})$ is the geometrical distance between the vertices $v_{i}$ and $v_{k}$. 
The distance $d(v'_{j},v'_{l})$ is defined similarly. When $d_t=0$, a restrictive 
3D representation is modelled, penalizing in situations where the distances between 
atoms in corresponding graphs are not equal. The conflict graph in this case has the 
highest density. Its density reduces as we allow for the modelling of increasingly relaxed 3D representations 
by increasing $d_t$. We eventually reach a model that represents the 2D structure when $d_t$ 
is greater than any of the distances between atoms. After the conflict graph is built, a QUBO problem
formulation can be obtained as explained in Sections \ref{subsec:cokplex_QUBO} and \ref{Similarity_Measure}. 

In the following section, we present a case study applying our molecular similarity 
method to predict the mutagenicity of molecules.

\section{Case Study: Prediction of Mutagenicity}
\label{sec:Exp_result}

In this section, we use a supervised machine learning approach to build a novel model that predicts 
the mutagenicity of chemical molecules. The predictive model is based on the proposed QUBO-based 
molecular similarity method described in Section \ref{molecularsimilarity}. We then present some results, 
validating the performance of the predictive model using two real data sets and 
compare it with the conventional fingerprint-based approach.

\subsection{Predictive Model}
\label{subsec:predictive_model}

In the pharmaceutical industry, the liability of drugs needs to be assessed before commercialization. To that end, 
researchers attempt to estimate a number of desired properties associated with absorption, distribution, metabolism, 
excretion, and toxicity for collections of drugs by using their molecular structures. One of the properties related 
to toxicology is mutagenicity. Detecting mutagenicity is a crucial step in drug discovery because mutagenic chemicals 
may potentially be carcinogenic. That is, they may damage DNA, resulting in mutations that can cause 
cancer. Using preclinical experimental tests to predict the mutagenicity of a  molecule 
is significantly time-consuming and expensive. Therefore, developing robust \emph{in silico} predictive techniques 
is of considerable importance in decreasing the time between early development and launch phases in
the life cycle of a drug. 

Our predictive model applies the $\kappa$-nearest neighbours ($\kappa$-NN) statistical method as the 
classifier to assign a ``mutagen'' or ``non-mutagen'' label to a molecule. Although there are 
various classifiers in the machine learning literature, we use the $\kappa$-NN classifier because 
of its simplicity and its well-known high performance in the context of computational biology 
\cite{Slonim02,Pomeroy02}.  The $\kappa$-NN classifier measures 
the pairwise distance between an unknown molecule $i$ and a data set of molecules with known mutagenicity 
labels. It then selects the $\kappa$ closest molecules to molecule $i$ from the data set and counts the 
contributions of mutagenic and non-mutagenic neighbours. Finally, the label with the 
highest contribution is assigned to molecule $i$. The pairwise distance 
between molecule $i$ and molecule $j$ in the data set, denoted by $\mathcal{D}_{ij}$, 
is equal to $1-\mathcal{S}(G_i,G_j)$, where $G_i$ and $G_j$ are the corresponding graphs 
of molecules $i$ and $j$, respectively, and $\mathcal{S}(G_i, G_j)$ is the similarity measure between the two graphs, 
quantified using Equation \eqref{simEq}. The contribution of neighbour $j$ is measured by its weight being 
equal to $1/\mathcal{D}_{ij}$. Although various weight functions have been proposed for the $\kappa$-NN classifier, our early 
experimentation showed that using the inverse of the distance as a weight function results in better performance.

To measure the similarity between two molecules, we need to assign 
values to the optional parameters, including the threshold distance between atoms ($d_t$), the number of hydrogen bonds 
associated to rings (RH), the bond order within a ring (RB), the formal charge (FC), and the degree of the nodes (DN). 
While the first parameter takes any positive real number, the last four are assigned a value of 1 if they are 
taken into account in constructing the conflict graph, or 0 otherwise. Furthermore, the 
co-$k$-plex relaxation parameter (i.e., $k=1$ or $k > 1$) should be determined. Different combinations of parameter values may 
result in different similarity values, changing the performance of the predictive model as a consequence. For 
each combination, we use a $k$-fold cross-validation technique to find the predictive model's performance. 
The combination yielding the best performance determines the best set of parameter values. 

The cross-validation is performed on a data set provided by Xu et al. \cite{Xu12} that contains 7617 molecules, 4252 of them mutagenic and 
3365 non-mutagenic. The data set is partitioned into five folds, where, initially, 
the first fold is reserved as the validation set and the other folds are used as the training set. 
To assign a ``mutagen'' or ``non-mutagen'' label to each molecule in the validation set, we use the 3-NN classifier. The performance 
of the classifier is then evaluated using four performance metrics, defined below. This procedure is 
repeated five times, each fold acting once as a validation set. Finally, the performance of the predictive model 
is equal to the average performance of the classifier over five folds. The number of folds and neighbours are chosen as 
five and three, respectively, because our early experiments showed that 
different values of folds and neighbours do not have a significant effect on the performance of the classifier. 

The four performance metrics used are sensitivity, specificity, accuracy, and precision. 
They are based on counting four numbers in the validation set: true positives ($TP$), true negatives ($TN$), 
false positives ($FP$), and false negatives ($FN$). Sensitivity (true positive rate) is equal to 
$TP/(TP + FN)$, the conditional probability that a 
molecule is labelled ``mutagen'' given it truly is mutagenic. Specificity (true negative rate) equals 
$TN$/($TN + FP$), the conditional probability that a molecule is labelled as ``non-mutagen'' given it truly 
 is non-mutagenic.  Accuracy is equal to ($TP + TN$)/($TP + FP + TN + FN$), the 
probability of predicting a molecule's label correctly. Precision is the likelihood of 
labelling a truly mutagenic molecule ``mutagen'', and is equal to $TP$/($TP + FP$). To compare predictive 
models, we first consider the accuracy metric. However, due to the accuracy 
paradox, a high accuracy value does not guarantee that a classifier is good. A good classifier not 
only has high accuracy, but also reasonable performance in terms of the other three metrics. 
Furthermore, since sensitivity is negatively correlated with specificity and precision, 
a good classifier is the one with balanced performance in terms of all metrics.

Before discussing the results, we provide an example to show how the similarity value between two molecules 
changes as the parameter values change.

\begin{figure}[h]
\centering
\includegraphics[scale=0.6]{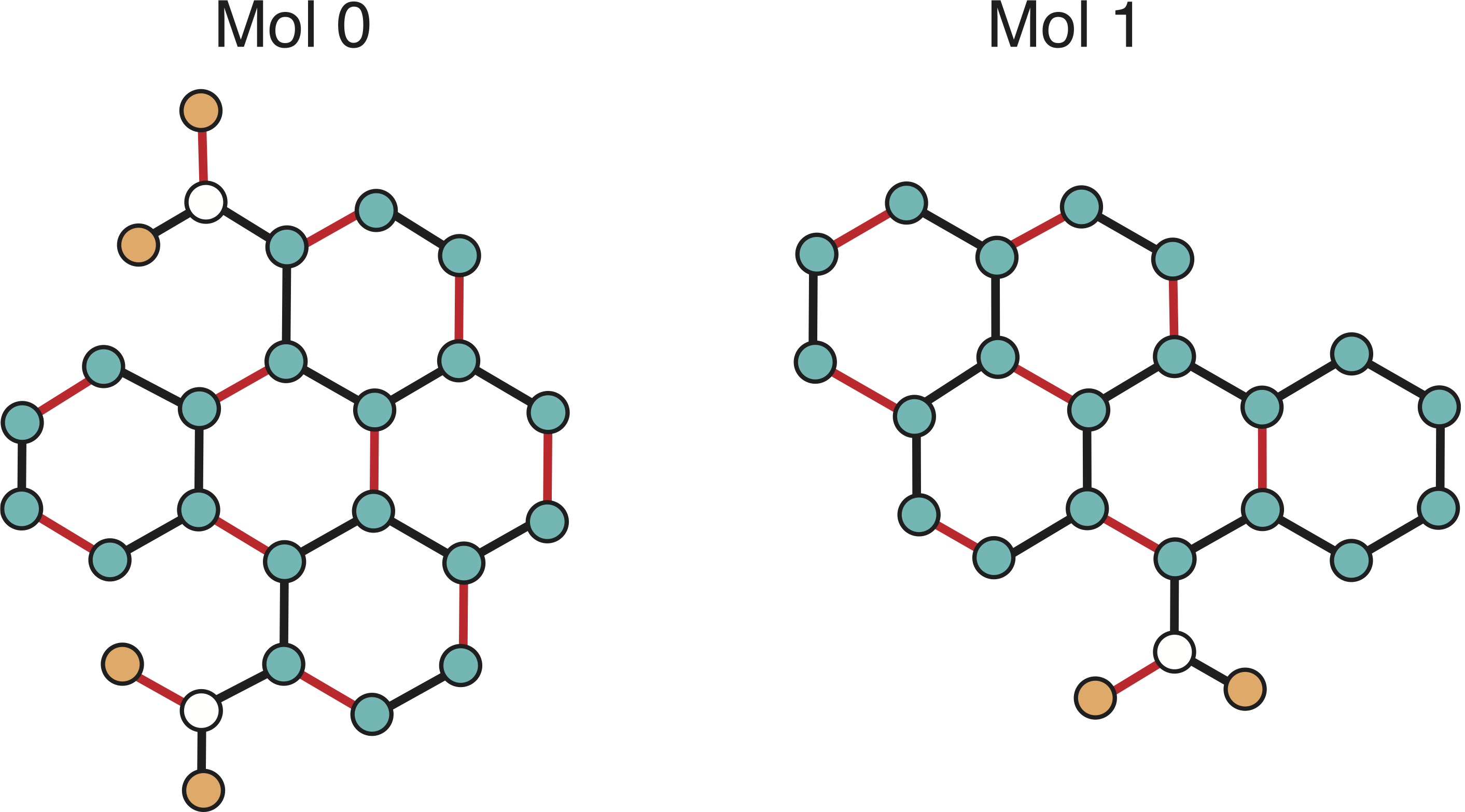}
\caption{The molecular representations of molecules 0 and 1.}
\label{fig:mol_exap}
\end{figure}

\paragraph{Illustrative Example} We measure the similarity between two molecules, ``1,8-Dinitrobenzo[e]pyrene" (Mol 0) and 
``6-Nitro-7,8,9,10-tetrahydrobenzo[pqr]tetraphene" (Mol 1), for different parameter values. The 
molecules are represented in Figure \ref{fig:mol_exap}. 

Table \ref{table:illustrative_example} shows the conflict graphs and the similarity values 
of Mol~0 and Mol~1 for different parameter values when $\delta=0.5$. The black, green, 
and blue edges are due to bijection, distance, and edge label conflicts, respectively. 

\begin{table}[h]
\small
\centering
\begin{tabular}{|c|c|c||c|c|} \hline
\multirow{1}{*}{Optional}  &  \multicolumn{2}{|c||}{$d_t=0$} &  \multicolumn{2}{|c|}{$d_t=1.5$} \\ \cline{2-3} \cline{2-5}
		  Parameters	             & \multicolumn{1}{|c}{$k=1$} & \multicolumn{1}{|c||}{$k=4$} & \multicolumn{1}{|c}{$k=1$} & \multicolumn{1}{|c|}{$k=4$} \\ \hline
FC = 1, DN = 1					     
                                                        & \raisebox{-\totalheight}{\includegraphics[width=0.15\textwidth, height=25mm]{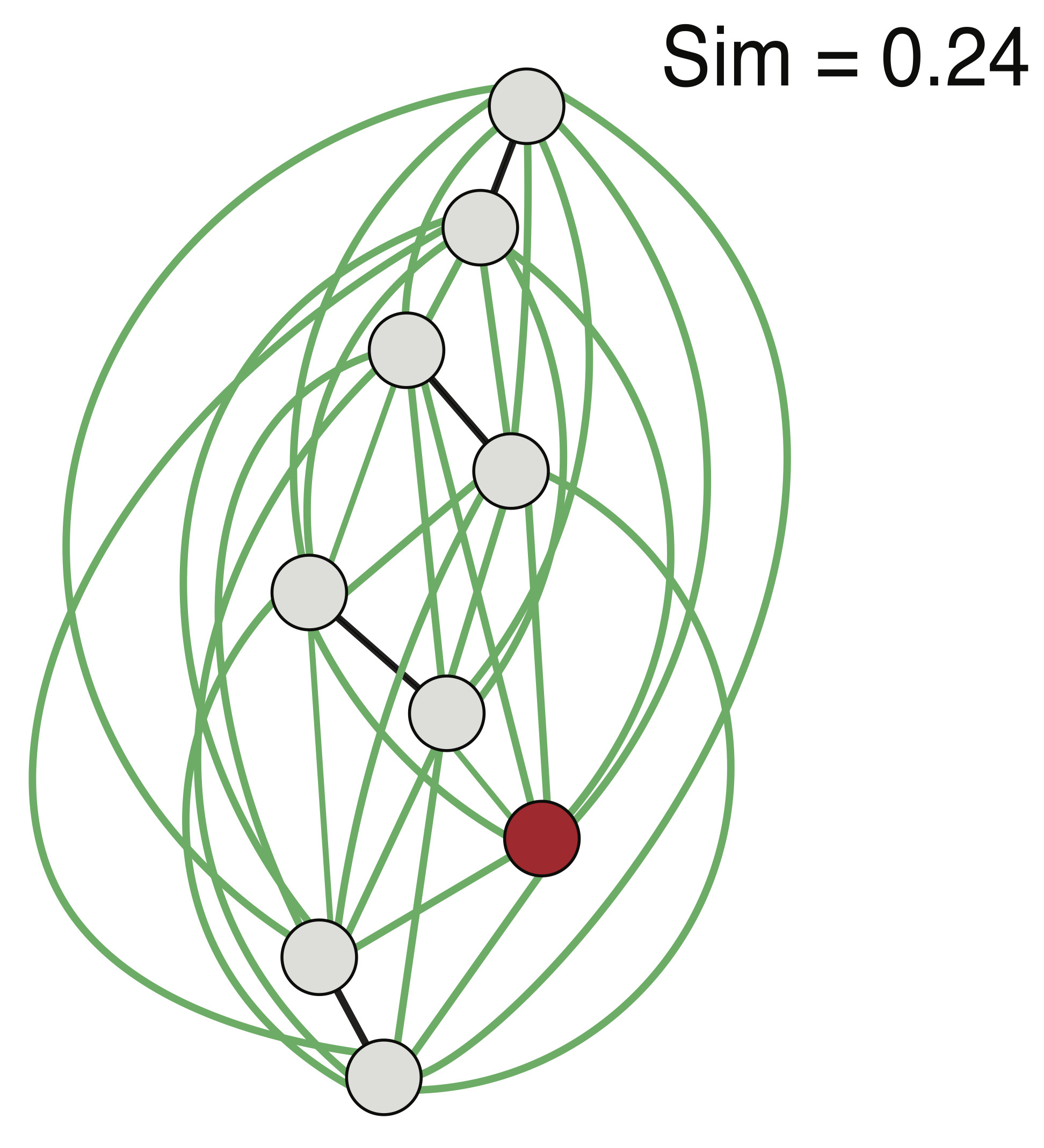}}             
						     &  \raisebox{-\totalheight}{\includegraphics[width=0.15\textwidth, height=25mm]{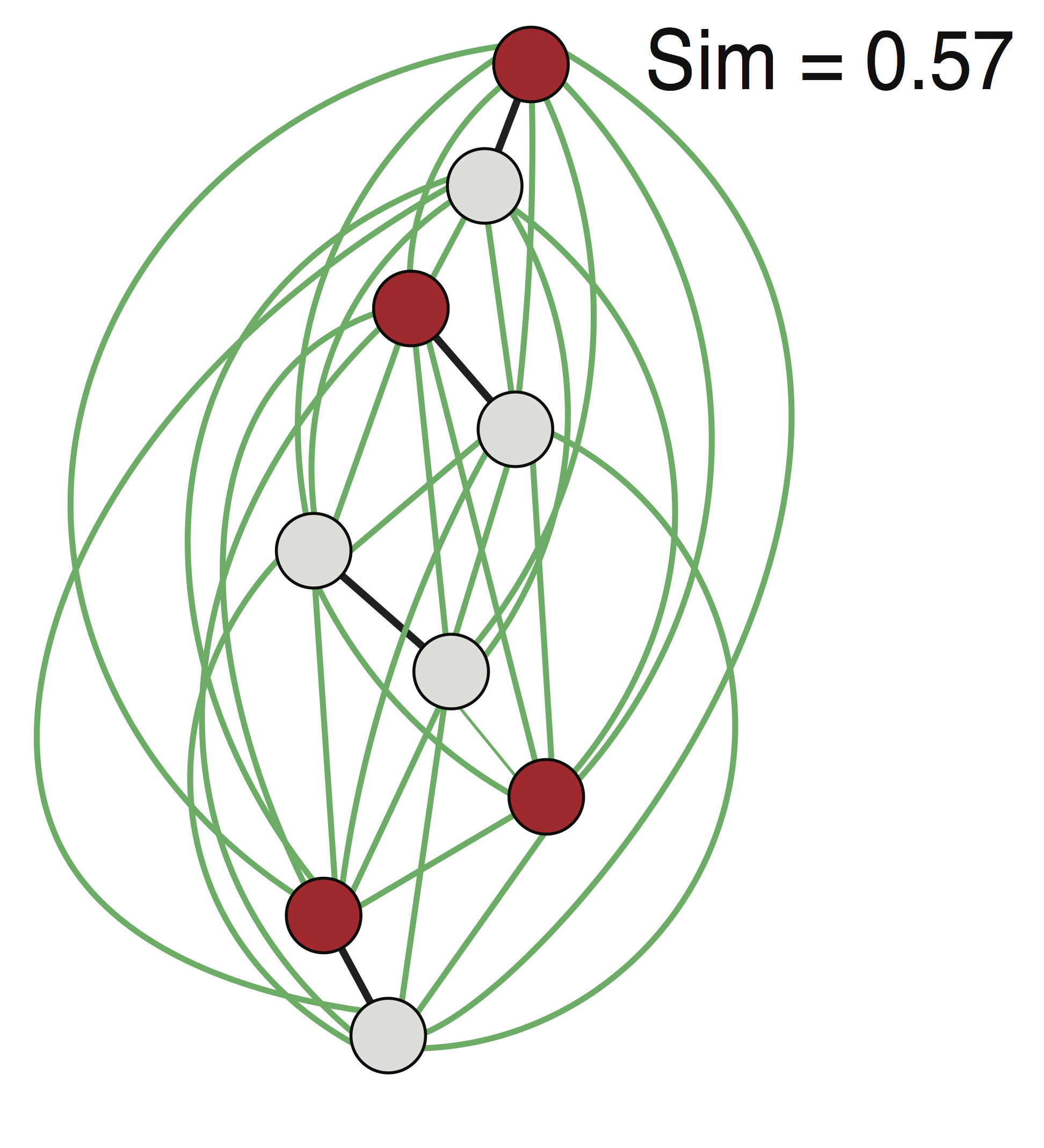}}                  
						     &  \raisebox{-\totalheight}{\includegraphics[width=0.15\textwidth, height=25mm]{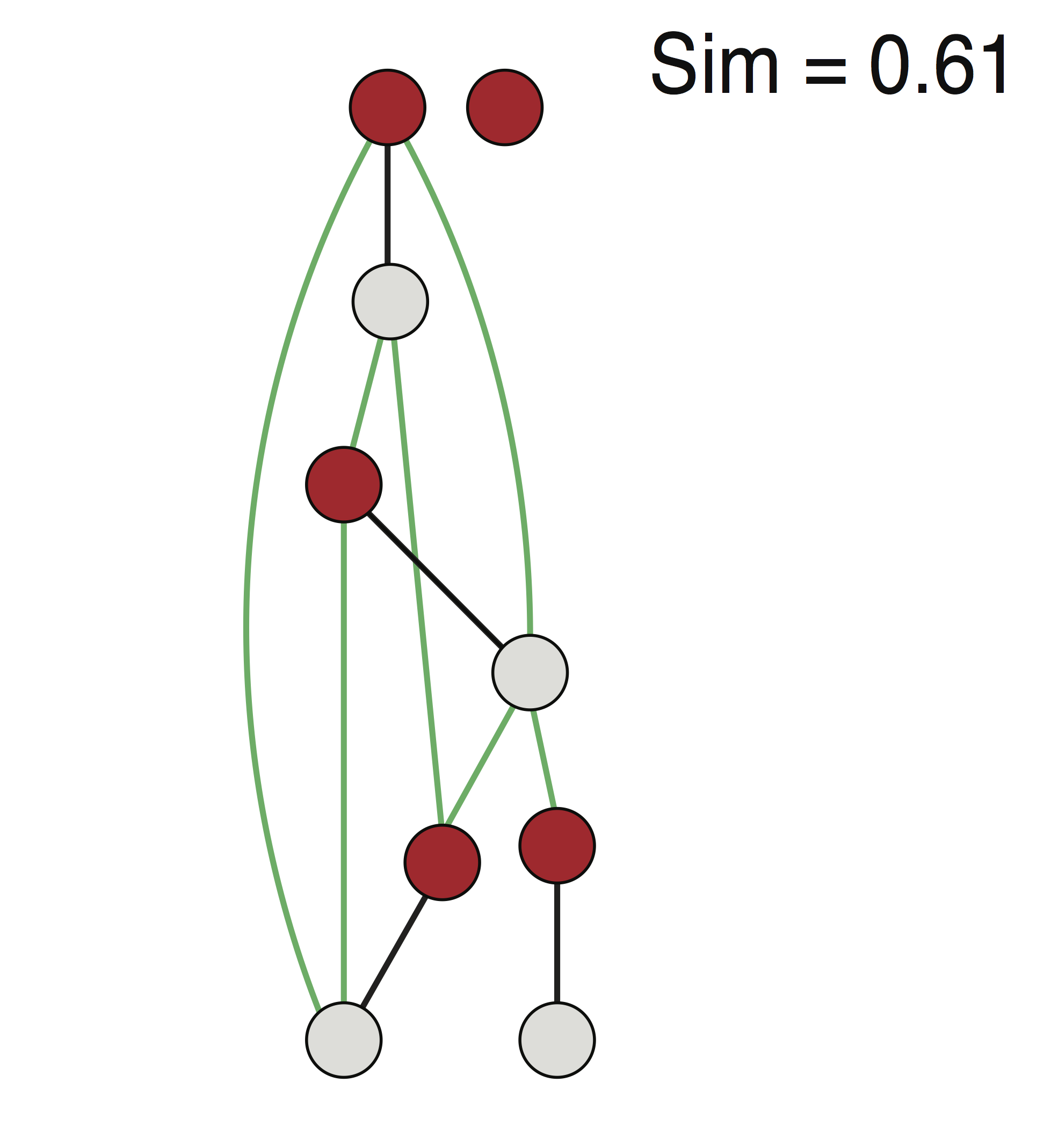}}         
						     &  \raisebox{-\totalheight}{\includegraphics[width=0.15\textwidth, height=25mm]{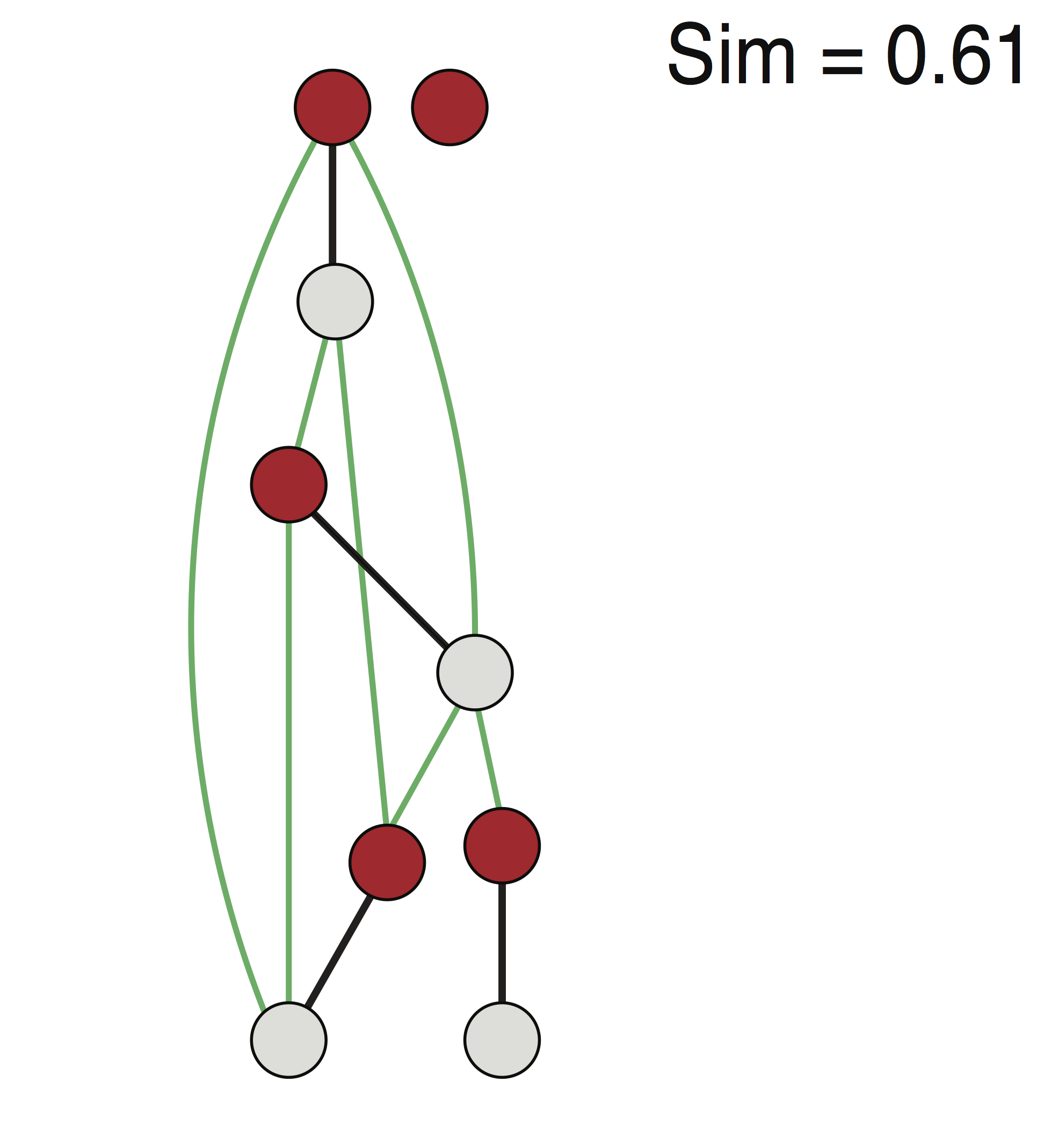}}               \\ \hline
RH = 1, FC = 1						     
						     &  \raisebox{-\totalheight}{\includegraphics[width=0.125\textwidth, height=25mm]{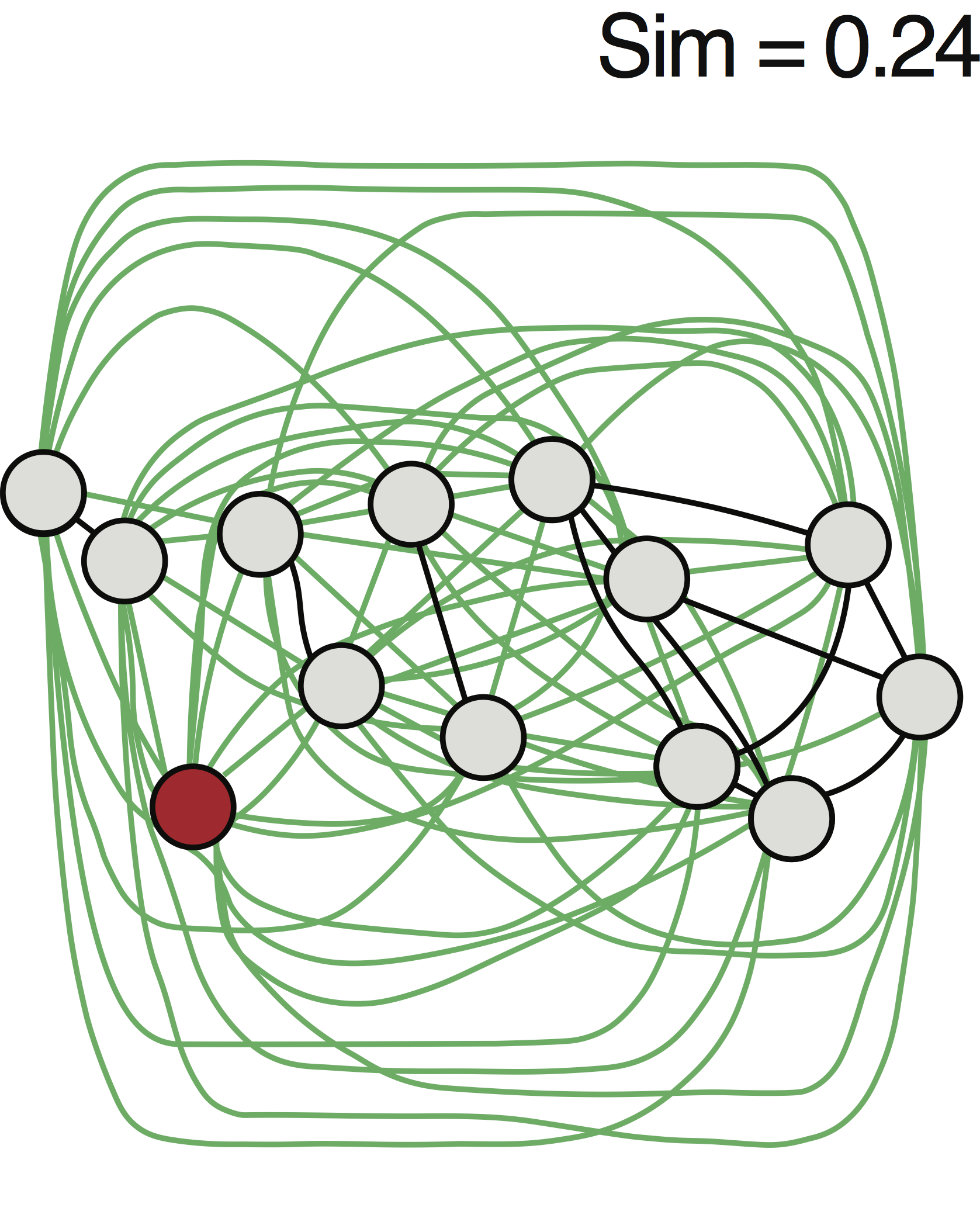}}              
						     &  \raisebox{-\totalheight}{\includegraphics[width=0.125\textwidth,height=25mm]{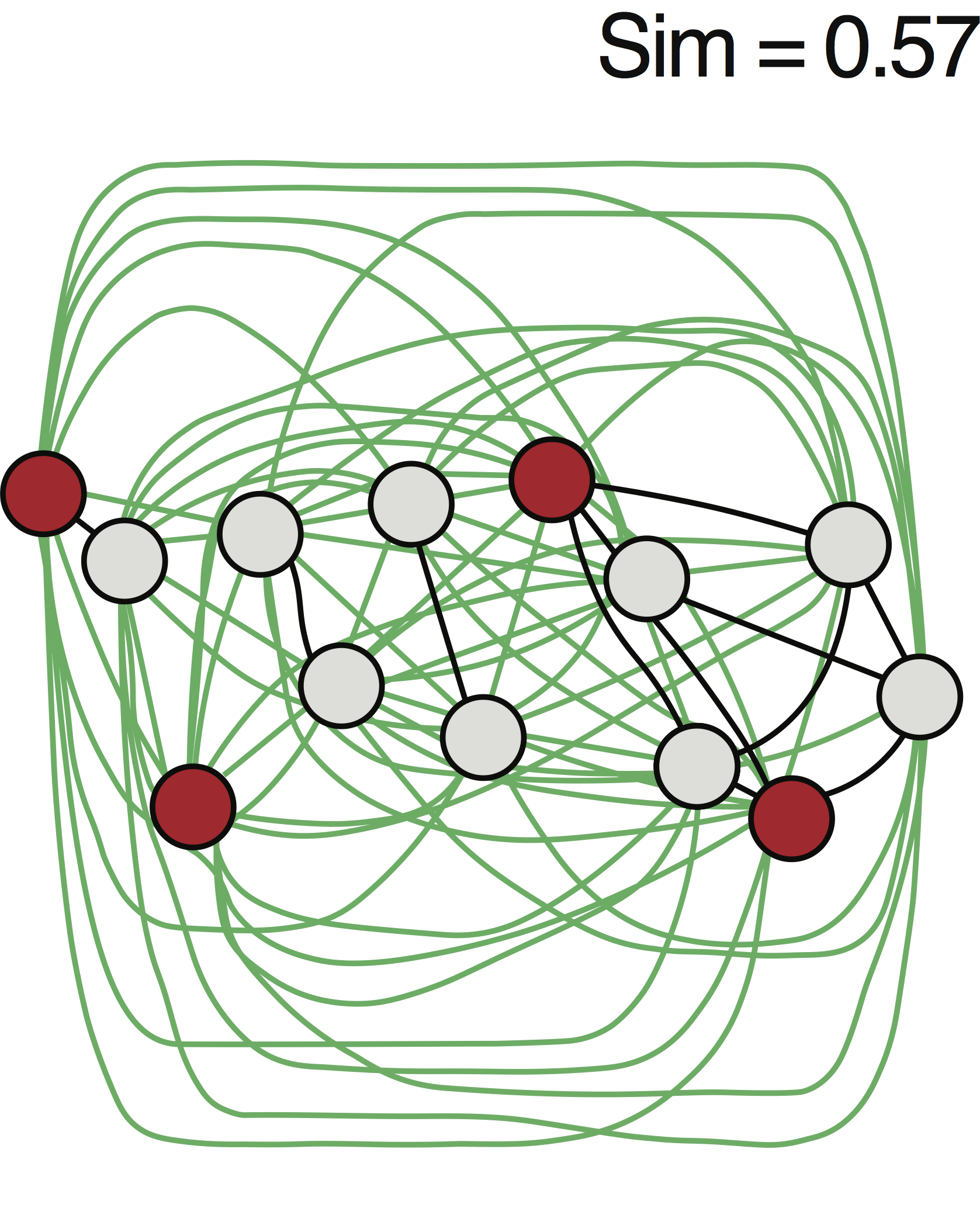}}             
						     &  \raisebox{-\totalheight}{\includegraphics[width=0.15\textwidth, height=25mm]{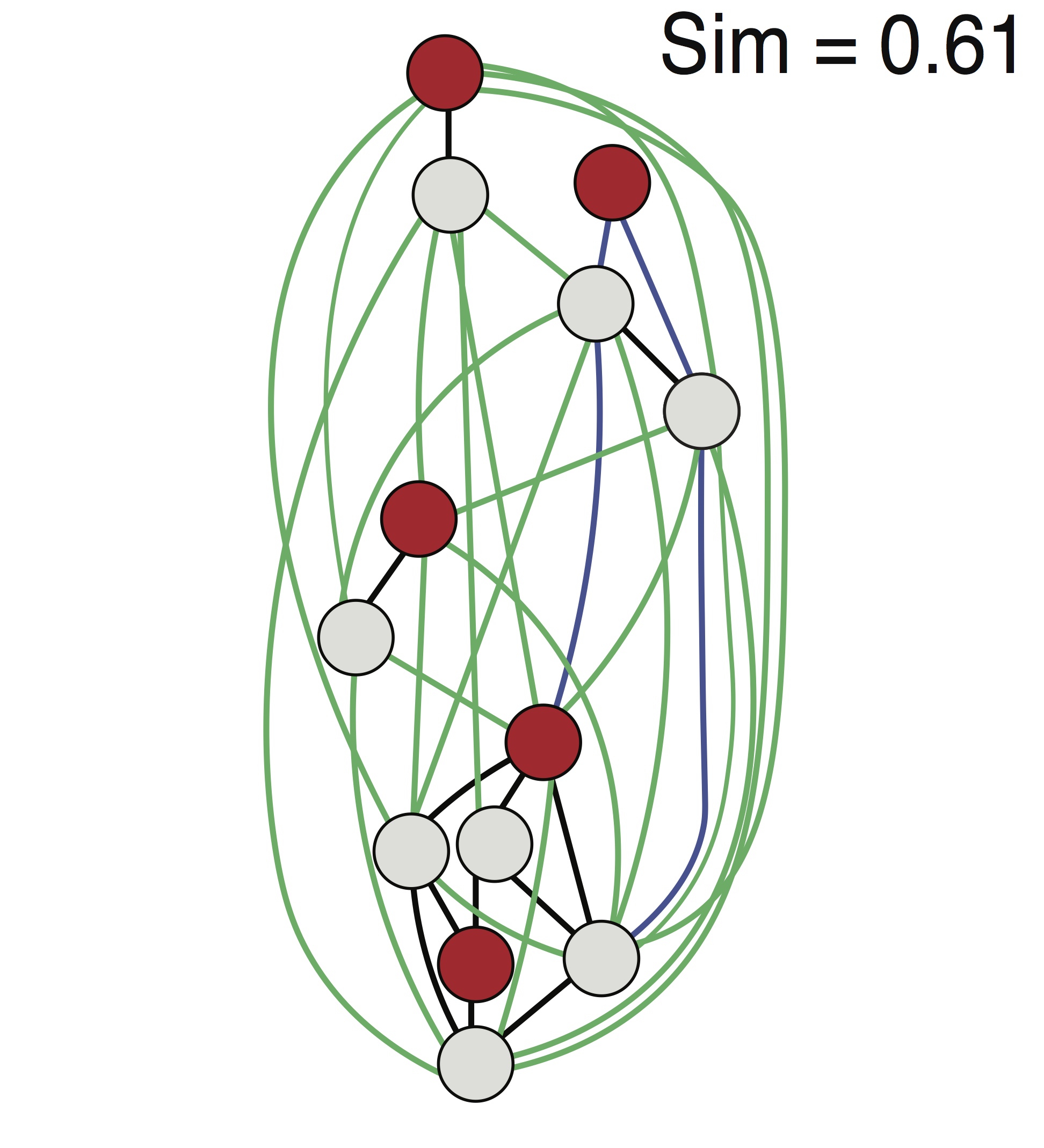}}             
						     &  \raisebox{-\totalheight}{\includegraphics[width=0.15\textwidth, height=25mm]{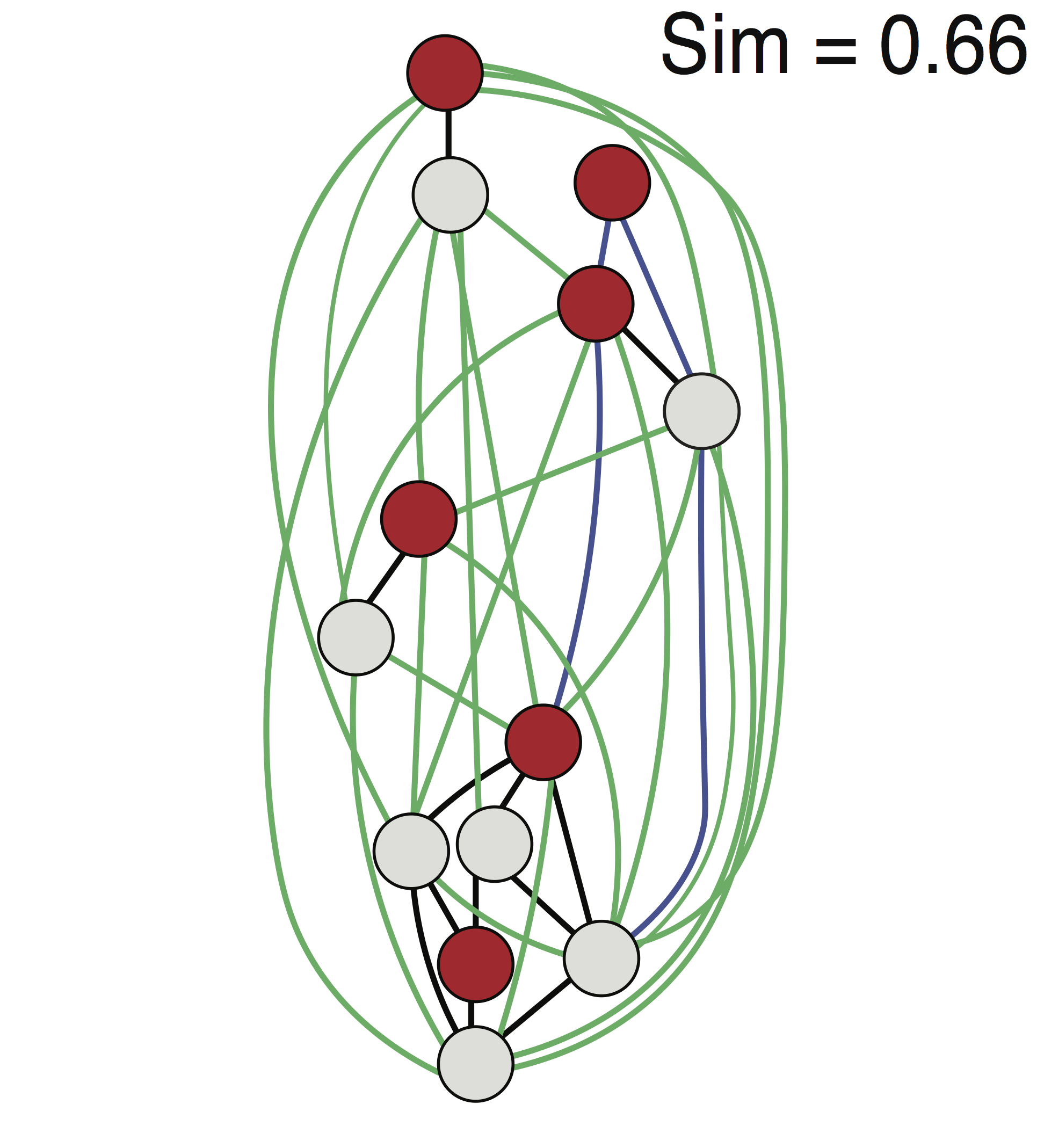}}       	 \\ \hline
\end{tabular}						     
\caption{The conflict graphs and similarity values of Mol 0 and Mol 1 for different parameter values when $\delta=0.5$.}
\label{table:illustrative_example}
\end{table}
 
As expected, the conflict graphs and similarity values change for different 
values of optional parameters, distance thresholds, and co-$k$-plex relaxations. 
More specifically, for given optional parameters, the conflict graphs are denser for 
lower distances and the similarity values increase or remain the same  as the similarity 
criteria become more relaxed by increasing either $d_t$ or $k$.
  
\subsection{Results and Discussion}
\label{subsec:results_discussion}
 
 In this section, we use the training data set of 7617 molecules to find the best set of 
 parameters, then validate the predictive model performance on two test sets, 
 and finally compare the predictive model with the fingerprint-based approach. To find the best set 
 of parameters, we investigate how different combinations of the optional molecular labels 
 RH, RB, FC, and DN change the performance metrics. We then report on the effect 
 of different distance threshold and co-$k$-plex relaxation values, $d_t$ and $k$, respectively. 
  
 \vspace{-0.5cm} 
 \paragraph{Molecular Feature Selection} As discussed earlier, there are four optional molecular 
 features---RH, RB, FC, and DN---that can be considered in 16 different layouts to build the conflict graph. 
 Each layout setting results in a different conflict graph size (see Table \ref{table:illustrative_example}). 
 In this paper, we use an exhaustive solver to solve the QUBO problem formulations of conflict graphs 
 whose number of vertices is less than or equal to twenty. The molecules that have a larger conflict 
 graph are excluded from the cross-validation. Therefore, the number of molecule pairs used in each replication 
 of the cross-validation differs for each of the different layouts. To make a fair comparison of the effect of different 
 molecular features, we have excluded the pairs whose similarity cannot be quantified in at least one of the layout settings 
 from the training sets of all layouts. In other words, we have considered the same molecule pairs in all layouts, 
 which we refer to as a ``reduced pairs set".
 
\begin{figure}[h]
\centering
\includegraphics[trim=40mm 0mm 40mm 0mm, width=11cm]{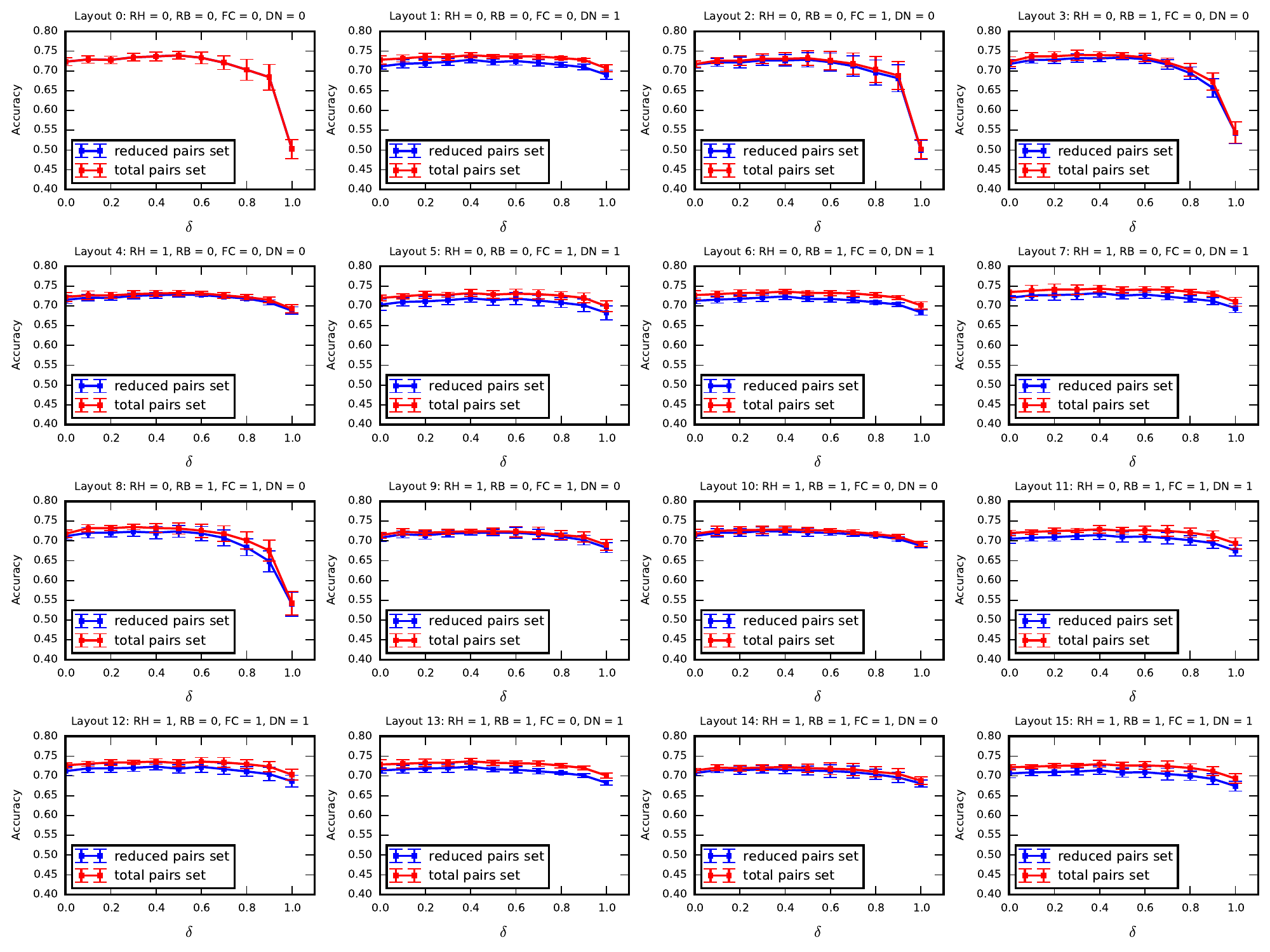}
\caption{Accuracy of 16 different layouts of molecular features with $d_t=1.5$ and $k=1$ in two cases: 
total number of pairs (total pairs set) and reduced number of pairs (reduced pairs set) solved.} 
\label{figure:accuracy_1.5}
\end{figure}

Figure \ref{figure:accuracy_1.5} illustrates the mean accuracy of our 3-NN classifier 
 with $k=1$ and $d_t=1.5$, along with error bars, for different layouts, considering 
 the total number of pairs and reduced number of pairs solved. As shown, Layout~0, where none
of the molecular features are taken into account, has 
the lowest number of pairs solved (the two accuracy plots are the same). Since 
excluding the molecular features increases the number of matching vertices in the conflict graph, 
it is more likely that there are more molecule pairs whose conflict graphs have more than 
twenty vertices in Layout 0 and their QUBO problem formulations cannot be solved. 
Figure \ref{figure:accuracy_1.5} also shows that for the rest of the layouts, the accuracy is greater when the total 
number of pairs solved is considered. 
 
We can further observe that when reduced pairs (blue plots) are considered, the highest accuracy achieved 
across all layouts is between $[0.71, 0.74]$, with Layout 0 having the highest accuracy. When the total number of pairs solved (red plots) are 
considered, the highest accuracy for all layouts slightly increases. The highest accuracy obtained across all layouts in this case is 
between $[0.72, 0.74]$, where Layout 7 has the highest accuracy. It is worth mentioning that Layouts 0, 1, 3, 8, 12, and 13 
also have a mean accuracy of 0.74. However, we report Layout 7 as being the best since it has a slightly higher 
accuracy---it is $0.744$ while the others are between $0.735$ and $0.740$. 
Given that the standard error of the accuracy is almost $1\%$ and the difference between the 
highest accuracy in Layouts 0 and 7 is 0.004, it is reasonable to consider both Layouts 0 (RH = 0, RB = 0, FC = 0, DN = 0) 
and 7 (RH = 1, RB = 0, FC = 0, DN = 1) as the best layouts for investigating the impact of the other parameters.
 
Figure \ref{figure:accuracy_1.5} also illustrates that the accuracy of all layouts, either with a reduced pairs set or 
total pairs set, is robust with respect to $\delta \in [0.3, 0.5]$. Recall that different values of $\delta$  
in Equation \eqref{simEq} yield different similarity values, affecting the performance of the 
classifier as a consequence. The robustness of the accuracy across a range 
of $\delta$ values affords us the advantage of being able to choose $\delta$ such that the classifier has 
reasonable performance with respect to the other three metrics---precision, sensitivity, and specificity. 
Table~\ref{tabel:layout07_pre_sen_spe} shows the four metrics for Layouts 0 and 7, 
where $\delta \in [0.3, 0.5]$. As expected, specificity and 
precision are negatively correlated with sensitivity. We could therefore choose 
$\delta$ to balance for the metric that is most relevant to the context of our 
classifier. In our predictive model of mutagenicity, it is more desirable to have a high 
sensitivity (i.e., most of the mutagenic molecules are identified) at a potential loss of precision and 
specificity where some of the non-mutagenic molecules are retrieved as mutagenic. 
Given Table \ref{tabel:layout07_pre_sen_spe}, setting $\delta=0.4$ seems reasonable in the context of
our case study.

\begin{table}[h!]
 \centering
\begin{tabular}{c|cccc|cccc} 
\multirow{2}{*}{$\delta$} & \multicolumn{4}{c}{Layout 0} & \multicolumn{4}{|c}{Layout 7} \\ \cline{2-9}
          	 & Accuracy        & Precision 	& Sensitivity         & Specificity        & Accuracy  & Precision  	& Sensitivity 		& Specificity \\ \hline
0.3          &    0.73             &  0.76        	&   0.78                &   0.68               &  0.74         & 0.76 		& 0.80                   & 0.68   \\
0.4          &    0.74             &  0.76  		&   0.77                &   0.70               &  0.74         & 0.76 		& 0.80                   & 0.68 \\
0.5          &    0.74             &  0.77    		&   0.76                &   0.72               &  0.74         & 0.76		& 0.79                   & 0.69 \\ 
\end{tabular}					     
\caption{Accuracy, precision, sensitivity, and specificity for Layouts 0 and 7 and different $\delta$ values.}
\label{tabel:layout07_pre_sen_spe}
\end{table}
 
\begin{figure}[h]
\centering
\includegraphics[trim=10mm 97mm 80mm 0mm, width=11cm]{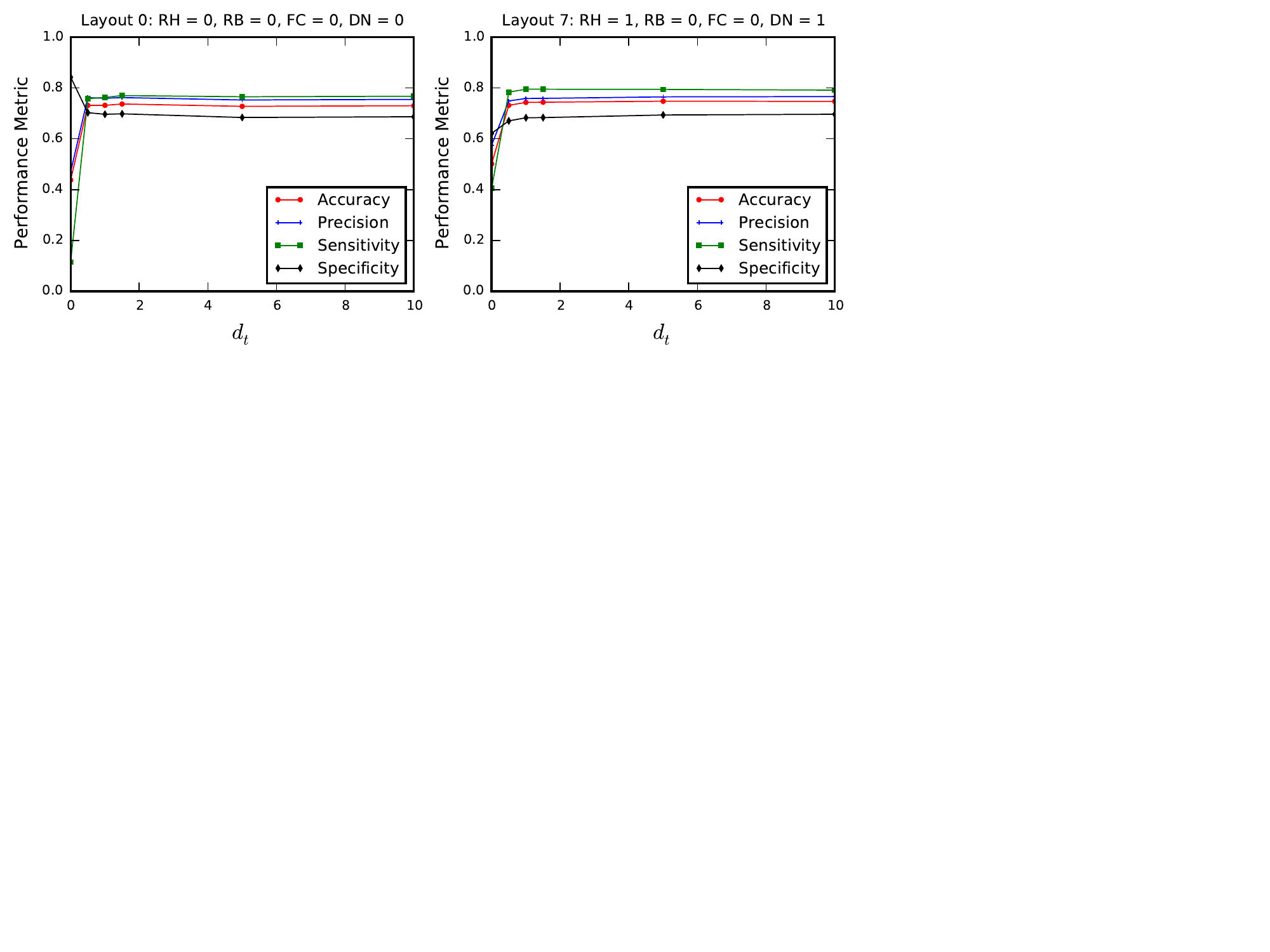}
\caption{Accuracy, precision, sensitivity, and specificity for Layouts 0 and 7 and different $d_t$ values when $k=1$ and 
$\delta=0.4$.}
\label{figure:layout7_distance}
\end{figure}

 \paragraph{Distance Threshold Parameter} As mentioned in Section \ref{graphbasedmolecularsimilarity}, the distance 
 threshold parameter $d_t$ allows some information regarding the 3D structure of the molecules to be
 incorporated into the conflict graph. The number of edges in the conflict graph of two molecules decreases as 
 $d_t$ increases. Therefore, the size of the maximum weighted independent set 
 of the conflict graph might increase, capturing more similarities between molecules. 
 However, it is clear that there is an upper bound on $d_t$ where the conflict graph 
 does not change. Therefore, the similarity values and consequently the performance of the 
 classifier remain constant. Figure \ref{figure:layout7_distance} illustrates the four performance 
 metrics of the classifier for Layouts 0 and 7 and six different values of $d_t$ 
 ($0$, $0.5$, $1$, $1.5$, $5$, and $10$) when $k=1$, $\delta=0.4$. We see that the classifier performs poorly at $d_t=0$, while 
 its performance substantially improves for $d_t=0.5$ and then 
 remains almost the same for higher values of $d_t$. 
 
 Intuitively, we would expect the accuracy for a given layout, $\delta$, and $k$ to be a concave and 
 monotonic function of $d_t$. The accuracy increases as $d_t$ increases, but it is bounded above since 
 the similarity values between molecules would eventually be invariant at higher $d_t$ values.
 
\paragraph{Co-$\emph{k}$-plex Relaxation} At higher values of $k$, we allow more conflicts to be 
present in the independent set of the conflict graph. As a result, the similarity value between 
two molecules increases. Table \ref{table:layout07_cokplex} shows the accuracy of Layouts 0 and 7 
for different $k$ and $d_t$ values, where $\delta=0.4$. The highest accuracy value is 0.75 for 
Layout 7, $k=3$, and $d_t=1.5$.

\begin{table}[h!]
 \centering
\begin{tabular}{c|ccccc|ccccc}
\multirow{2}{*}{$d_t$} & \multicolumn{5}{c}{Layout 0} &   \multicolumn{5}{|c}{Layout 7} \\ \cline{2-11}
                  & $k=1$    & $k=2$   & $k=3$     & $k=4$    & $k=5$ & $k=1$    & $k=2$   & $k=3$     & $k=4$    & $k=5$ \\ \hline  
0                &    0.44    &    0.48   & 0.72        & 0.65       &  0.65   &   0.50     &   0.61    &   0.67     &   0.69     &  0.71        \\ 
0.5             &    0.73    &    0.72   & 0.71        & 0.70       &  0.71   &   0.73     &  0.74     &   0.74     &  0.74      & 0.74         \\ 
1                &    0.73    &    0.73   & 0.72        & 0.71       &  0.71   &    0.74    &  0.74     &  0.74      & 0.74       & 0.74          \\ 
1.5             &    0.74    &    0.73   & 0.71        & 0.71       &   0.70  &  0.74      &  0.74     &  \textbf{0.75}      & 0.74       & 0.74 \\  
\end{tabular}					     
\caption{Accuracy of Layouts 0 and 7 for different $k$ and $d_t$ values, where $\delta=0.4$.}
\label{table:layout07_cokplex}
\end{table}

Table \ref{table:layout07_cokplex} also shows that when $d_t=0$, the accuracy substantially improves 
at higher values of $k$ for both Layouts 0 and 7. It changes from a useless 
(i.e., randomly assigning mutagenicity labels to molecules) to a useful classifier. 
For distance $d_t=0$, the accuracy in Layout 7 has a strictly increasing trend, whereas the accuracy 
in Layout 0 reaches its highest value at $k=3$ and then decreases. We can further observe 
that for $d_t >0$, the difference in accuracy diminishes as $k$ increases. 
This observation can be justified since we know that the number of conflict edges decreases for higher 
values of $d_t$ and that higher values of $k$ are more beneficial when the conflict graph is denser. 
To conclude, we expect that the accuracy would be a non-monotonic function of $k$ for a given $d_t$. 
That is, there would be eventually a $k_{\footnotesize{\mbox{max}}}$ such that the performance 
of the classifier for all $k > k_{\footnotesize{\mbox{max}}}$ is lower than its performance 
at $k_{\footnotesize{\mbox{max}}}$, ultimately reaching a constant limit.

\paragraph{External Validation and Benchmarking} Our investigation indicates that the set of parameter values including Layout 7,
$d_t=1.5$, $k=3$, and $\delta=0.4$ obtains the highest accuracy of 0.75 $\pm$ 0.01. 
We use these parameter values to validate our QUBO-based maximum weighted co-$k$-plex relaxation 
model on two external test sets. The first test set is a balanced set provided by Xu et al. \cite{Xu12}, 
containing 234 molecules with the same number of mutagens and non-mutagens. The 
second test set is built using the sets given by Hansen et al. \cite{Hansen09} and Xue et al. \cite{Xu12}. 
Hansen et al. constructed a data set comprising 6512 molecules, which we compare 
with our training set of 7617 molecules, excluding the molecules 
that are common to both. Combining the 106 molecules that are not in our training set with 731 molecules of 
another set given by Xue et al. forms a new set of molecules. The second test set contains 300 molecules 
randomly chosen from the new set where half are mutagenic and half are non-mutagenic. 

The performance of our similarity model on both test sets is compared with a fingerprint-based approach. We use the MACCS fingerprint, a vector of 166 bits representing molecular features. Table \ref{table:test_fingerprint} 
shows the four performance metrics of the 3-NN classifier on both test sets, where the QUBO-based maximum 
weighted co-$3$-plex, the QUBO-based maximum weighted co-$1$-plex, and the fingerprint-based approaches are used. In Table \ref{table:test_fingerprint}, 
set 1 and set 2 refer to the first test set and the second test set, respectively.  The fingerprint-based approach measures 
the similarity between two molecules as ``$\mathrm{similarity} =1- \mathrm{distance}$". We use the Euclidean distance 
between two molecular vectors. For a fair comparison, 
we exclude the molecule pairs whose similarity values are not found in Layout 7 from the training set of 
the fingerprint-based method. As shown, both QUBO-based maximum weighted co-$k$-plex methods have better performance than 
the fingerprint-based method in three metrics---accuracy, precision, and specificity---in both test sets. 
Although the fingerprint-based method has a higher sensitivity in both sets, its low specificity value implies that 
the majority of non-mutagenic molecules are blindly labelled as mutagens, increasing the sensitivity. 
It is worth mentioning that our graph-based maximum weighted co-$k$-plex relaxation similarity measures not only 
yield a higher classification quality, but also provide complete information 
on the common molecular substructures while the fingerprint-based measure reports only one value without providing any insights. 

Table \ref{table:test_fingerprint} also shows the superiority of the maximum weighted co-$3$-plex relaxation method over 
the maximum weighted co-$1$-plex relaxation. The higher accuracy of the maximum weighted co-$3$-plex relaxation 
method provides evidence that accounting for the noisy data results in a more accurate prediction of mutagenicity.  

\begin{table}
\centering
\begin{tabular}{l|l|c|c|c} 
\multirow{3}{*}{Test Set}				& &  \multicolumn{3}{|c}{Method} \\ \cline{3-5}
							& \multicolumn{1}{c|}{Performance } 	    &  QUBO-based Maximum          & QUBO-based Maximum                & MACCS \\
							&      \multicolumn{1}{c|}{Metric}            &  Weighted Co-$3$-plex           &  Weighted Co-$1$-plex & Fingerprint  \\ \hline
\multirow{4}{*}{Set 1}      				&  Accuracy	   & 		\textbf{0.81}				  &	0.80					     &  0.76 \\ 
									& Precision	   &		0.79				  &	0.77					     & 0.69 \\ 
       									& Sensitivity	   &		0.84				  &	0.85					     & 0.95 \\ 
									& Specificity	   &		0.78				  &	0.75					     & 0.58 \\ \hline
\multirow{4}{*}{Set 2}      				&  Accuracy	   & 		\textbf{0.80}				  &	0.79					     & 0.77 \\ 
									& Precision	   &		0.78				  &	0.77					     & 0.70 \\ 
       									& Sensitivity	   &		0.83				  &	0.85					     & 0.95 \\ 
									& Specificity	   &		0.77				  &	0.74					     & 0.60 \\ 
\end{tabular}															
\caption{Accuracy, precision, sensitivity, and specificity of the 3-NN classifier, where 
the QUBO-based maximum weighted co-$3$-plex, the QUBO-based maximum weighted co-$1$-plex, and the MACCS fingerprint-based approaches are used.}
\label{table:test_fingerprint}
\end{table}

\section{Conclusions}
\label{sec:conclusion}

In this paper, we address the problem of 
measuring similarity among graphs with noisy data whose graph representations are partially accurate in the 
context of predicting molecular mutagenicity. To account for data distortion, we develop a novel QUBO 
problem formulation for the maximum weighted co-$k$-plex problem to measure the similarity 
of graphs, relaxing the requirement that the subgraph matching be exact. The relaxation 
allows for the existence of dissimilarities among graphs up to a predefined level. 
The QUBO-based similarity measure can also be extended to the problem of measuring similarity among multiple 
graphs.

The QUBO-based similarity measure is then used to quantify the similarity of molecules. 
We discusse efficient approaches to the abstraction of molecules as labelled graphs 
and build their corresponding conflict graphs, incorporating various molecular features, including 
3D structural information. To assess the performance of the developed measure, two real-molecule data 
sets are used to predict mutagenicity, exploiting a machine learning approach. Our results demonstrate 
that relaxing the definition of similarity by setting $k=3$ yields the highest prediction accuracy.  
Investigation of the results further indicates that the accuracy of mutagenicity prediction is poor 
when the strict 3D representation is modelled. There is also a maximum value for the level of 
relaxation in the maximum weighted co-$k$-plex formulation where the prediction accuracy is the highest. Finally, 
comparison of the QUBO-based similarity measure with the existing fingerprint-based measure
in the literature shows the superiority of our measure.

Although the developed similarity measure can be solved by a quantum annealer, we use an exhaustive solver to find 
the optimal QUBO-based similarity measures. We will pursue the study of the potential speed-up of 
the quantum annealing solver on the maximum weighted co-$k$-plex problem in future work.

 \section*{Acknowledgements} 

We would like to thank the software development team at 1QB Information Technologies (1QBit) for supporting the development of our code, Hamed Karimi for useful discussions, Robyn Foerster for valuable support, Marko Bucyk for editorial help, and other researchers at 1QBit for helpful input and comments. This research was supported in part by 1QBit and the Mitacs Accelerate and Elevate programs.



\end{document}